\documentclass[pra,10pt,twocolumn,superscriptaddress]{revtex4}
\usepackage{amsmath}
\usepackage{latexsym}
\usepackage{amssymb}
\usepackage{graphics,epstopdf}
\usepackage{graphicx}
\usepackage[colorlinks=true, citecolor=blue, urlcolor=blue ]{hyperref}
\usepackage{float}
\usepackage{graphicx}
\usepackage{amsfonts}

\begin{document}

\title{Witnessing entanglement sequentially: Maximally entangled states are not special}

\author{Anindita Bera}
\affiliation{Department of Applied Mathematics, University of Calcutta, 92 A.P.C. Road, Kolkata 700 009, India}
\affiliation{Harish-Chandra Research Institute, HBNI, Chhatnag Road, Jhunsi, Allahabad 211 019, India}

\author{Shiladitya Mal}
\affiliation{Harish-Chandra Research Institute, HBNI, Chhatnag Road, Jhunsi, Allahabad 211 019, India}

\author{Aditi Sen(De)}
\affiliation{Harish-Chandra Research Institute, HBNI, Chhatnag Road, Jhunsi, Allahabad 211 019, India}

\author{Ujjwal Sen}
\affiliation{Harish-Chandra Research Institute, HBNI, Chhatnag Road, Jhunsi, Allahabad 211 019, India}

\begin{abstract}
We investigate sharing of bipartite entanglement in a scenario where half of an entangled pair is possessed and projectively measured by one observer, called Alice, while the other half is subjected to measurements performed sequentially, independently, and unsharply, by multiple observers, called Bobs. 
We find that there is a limit on the number of observers in this entanglement distribution scenario. In particular, for a two-qubit maximally entangled initial shared state, no more than twelve Bobs can detect entanglement with a single Alice for arbitrary -- possibly unequal -- sharpness parameters of the measurements by the Bobs. Moreover, the number of Bobs remains unaltered for a finite range of near-maximal pure initial entanglement, a feature that also occurs in the case of
equal sharpness parameters at the Bobs. 
Furthermore, we show that for non-maximally entangled shared pure states, the number of Bobs reduces with the amount of initial entanglement, providing a coarse-grained but operational measure of entanglement.
\end{abstract} 
 

\maketitle

\section{Introduction}

Entanglement of a compound quantum system can be seen as growing out of the fact that 
the best possible knowledge of an entire system is not contained in 
the best possible knowledge of its subparts, even for pure states~\cite{schro,epr1935}. This bizarre phenomenon marks a counter-classical nature of quantum correlation, and its role in the context of information theory can hardly be overemphasized. On the one hand, manifestation of entanglement leads to a paradigm shift of our understanding of physical laws by rejecting a local realistic description of nature -- the Bell theorem \cite{bell}. On the other hand, entanglement is the key resource for tasks which cannot be performed by classical resources \cite{horodecki}. 
Examples of such tasks include
quantum dense coding \cite{dens}, quantum teleportation \cite{tele}, and cryptography utilizing Bell's theorem \cite{ekert,Barrett-Hardy,Renner-wolf}. Entanglement has also been argued to play an important role in the reduction of classical communication complexity \cite{cc}, quantum computing \cite{knil,BriegelNATURE}, understanding of many-body phenomena \cite{ujjwal-adp,osterloh,maciej-book}, emergence of classicality \cite{zurek-chapter,zurek-rmp,markham03,Angelo05}, etc. 

As with other resources, such as energy and information, one would like to have a quantitative theory of entanglement providing specific rules of detection, manipulation, and quantification \cite{horodecki}. Determining whether a state is entangled is one of the basic tasks of quantum information. In principle, one can determine the full quantum state via  state tomography, and subsequently apply a criterion for detection of entanglement. 
Notable ones include partial transposition \cite{peres-ppt,Horodecki-PLA}, which is necessary and sufficient for identifying entanglement of two-qubit and qubit-qutrit states, majorization \cite{Marshallbook,mazor1,mazor2,mazor3}, realignment \cite{alignment1,alignment2,alignment3}, and covariance matrix criteria. From the perspective of experimentally ascertaining whether a state is entangled, entanglement witnesses (EWs) \cite{Woronowicz76,Kryszynski79,Choi82,Horodecki-PLA,lewen,terhal2,terhal,dagmar-wit,ew,wit2,review-Sreetama} play an important role,
since they may require only a few local measurements.

Detection of entanglement provides a qualitative answer to our quest for understanding entanglement. 
The next step is to quantify the same, and broadly there are 
two approaches to do so, viz. ``operational'' and ``distance-based''.
Distillable entanglement~\cite{distil,efor,rains99,plenio007} and entanglement of formation~\cite{efor} are examples of operational measures of entanglement, that are conceptualized 
via state preparation procedures, which in turn is related to quantum information tasks like 
quantum communication.
 In distance-based approaches, state functions like  
geometric measure of entanglement \cite{shimony95,barnum01,geoem,blasone08,aditiujjwal010} and relative entropy of entanglement \cite{shimony95,vedral-rippin97,vedral1619,vedral-RMP02} have been proposed. 
The basic and important property that all the measures satisfy is that they do not increase, on average, under local quantum operations and classical communication (LOCC).
They are in general hard to compute, even numerically, after the corresponding state has been 
reconstructed via state tomography, with the latter being experimentally costly.
Experimental proposals for direct estimation of entanglement measures have been proposed, but 
these still remain difficult with currently available technology \cite{phoro01,ekrt002,ekertalves02,phoroprl003}.

Undoubtedly, even partial preservation of entanglement in a shared state in spite of  a few cycles of local operations performed by the sharing parties can be important for information processing schemes in which entanglement is utilized as a resource. The question that we are going to address in this paper is exactly along these lines. One may be inclined to think that 
such a scenario 
is 
related to the concept of  ``monogamy'' of quantum correlations \cite{sanders-review,qic},
which explores the extent to which  entanglement and other quantum correlations -- unlike classical correlations -- cannot be shared arbitrarily between many parties.
More precisely, Coffman \emph{et al.}~\cite{ckw} showed that if
a system in possession of ``Alice" is already entangled with that of ``Bob", then Alice's system can have only a limited amount of entanglement with a third system in possession of ``Charu". 
In the context of violation of local realism, it was shown that for a tripartite state shared between Alice, Bob, and Charu, if Alice and Bob can demonstrate violation of the Clauser-Horne-Shimony-Holt
(CHSH) inequality \cite{chsh}, then Alice and Charu or Bob and Charu cannot \cite{Kaszlikowski01,tv,oliveira13}. Monogamy of entanglement has applications ranging from  quantum cryptography \cite{ekert,terhalIBM,aninditaCRYPTO} to phase detection in many-body systems \cite{Chandran07,dhar11,Allegra011,song013,rao013,wang014,zhang016,Sadhukhan16,stupido1}. 

An independent path for 
distribution of entanglement between several observers, 
known as
entanglement splitting, was introduced by Bru{\ss} \cite{brub}, where it was asked that whether it is possible for a party, possessing half of a pure bipartite quantum state, to transfer some of her/his entanglement with the other party to a third party. Specifically,
in the case of a singlet pair shared between Alice and Bob, the question is to find the extent to which Bob's entanglement with Alice can be shared ``symmetrically
and isotropically" with a third party, called ``Badal", so that she could teleport, imperfectly, a quantum
state to both of them. Thus a
channel bifurcation is created with one input side (Alice), and two output
sides (Bob and Badal). Note that Bob's action is local in the
sense that he does not act on Alice's side while it is not local with
respect to Badal.  It was shown that independent of the amount of entanglement of the initial state, it is always possible to split it between more than two objects and the splitting can go up to an  arbitrary number of objects. However,  to have a non-classical teleportation fidelity in the output, the initial entanglement has to be greater than some threshold value depending on number of channel bifurcations.
One should also mention here the work by Bu\v{z}ek \emph{et al.}~\cite{buzek}, which showed how entanglement of a pair can be locally broadcast, and the later work which found that the same cannot be performed if more than two pairs are required at the output \cite{som,cirac003}.

In a related but different scenario, Silva \emph{et al.}~\cite{sygp} explored  a new fundamental question in the domain of violation of local realism: Can the violation of local realism of an entangled pair be distributed among particles with multiple observers, that act sequentially and independently of each other? In this context, when Alice possesses half of an entangled pair and several Bobs measure sequentially and independently on the other half, it was shown \cite{sygp,sm} that not more than two observers can demonstrate violation of the CHSH inequality. See also \cite{Curchod17,Li-Guo}.  Quantum steering~\cite{schrodinger-steering,reidsteering,wiseman} of a single system by multiple observers has also been demonstrated recently \cite{sm2}, going beyond the monogamy restriction on steering \cite{mreid}. 


 In the present work, we inquire about the maximal number of observers, called Bobs, possessing half of an entangled pair and measuring sequentially and independently, who can detect entanglement,  while the other half is possessed by another observer, called Alice, who performs projective measurements. The success of sequential measurements in preserving entanglement depends on the fuzziness present in each measurement apparatus.
For a maximally entangled initially shared state of two spin-$1/2$ systems, we find that at most
twelve Bobs can detect entanglement with Alice provided the sharpness parameter of each measurement apparatus used by Bobs are allowed to be different. The result is in sharp contrast to the entanglement splitting \cite{brub} and sharing of violation of local realism \cite{sygp} scenarios. Interestingly, we observe that the maximum number of Bobs who can successfully detect entanglement after sequential and independent measurements remains unaltered, even when the  shared initial state is not maximally entangled. Specifically, the maximal number of Bobs remains invariant at the value twelve until the entanglement of the initial pure state goes below $0.942$ ebits. This result implies that for the protocol at hand, maximally entangled states do not possess any special status. For similar findings, see \cite{huelga1997,acin002latorre,MAU2003}. We also observe that the maximum number of Bobs, witnessing entanglement with a single Alice, decreases with the decrease of the entanglement content of the initially shared state. Therefore, the number of successful Bobs demonstrating entanglement detection in this scenario, turns out to be an operational, albeit coarse-grained, measure of entanglement. 
It may be mentioned here that quantification of entanglement from an operational perspective is an important task as it potentially has practical ramifications. 
If we assume that all the measurements performed by the Bobs are equally weak, the maximal number that can identify entanglement turns out to be five for a shared state having entanglement not less than $0.924$. 
The scenario of different sharpness parameters used by  different Bobs can
 appear when the Bobs are situated in different laboratories but have near-noiseless quantum channels
 between them. On the other hand, a plausible scenario where the Bobs use the same 
 sharpness parameter for their measurements is when they act in the same laboratory (and the same 
 apparatus) but at different times.

We arrange the paper in the following way. In Sec.~\ref{ent-wit-unsharp}, we briefly discuss about detection of entanglement through witness operators and about the unsharp measurement formalism. In Sec.~\ref{scenarip}, we describe the scenario that we consider in this paper of distribution of the resource state. Next, in Sec.~\ref{sharing}, we demonstrate our results, followed by concluding remarks in Sec.~\ref{conclude}.

\section{Gathering the tools}
\label{ent-wit-unsharp}
In this section, we briefly describe the idea of entanglement witnesses and unsharp measurements.

\subsection{Entanglement witnesses} 
\label{ent-wit}
An important problem in quantum information is the detection of entanglement in quantum state. Any (linear) observable which has at least one negative eigenvalue and has a non-negative average on all product states,  
can be used to detect entanglement.
These observables have been named as (linear) entanglement witnesses (EWs)\cite{Woronowicz76,Kryszynski79,Horodecki-PLA,lewen,terhal2,terhal,dagmar-wit,ew,wit2,review-Sreetama},
and provide an useful method for experimental detection of entanglement.
More precisely, an entanglement 
 witness is a Hermitian operator, denoted by $W$, that satisfies the following
\begin{eqnarray}
\label{EW_def}
\exists~ \text{at least one}~ \rho \notin {\cal S},~ \text{s.t. } \text{Tr}(W\rho) &<& 0 \nonumber \\
 \text{while}~ \forall \rho_s \in {\cal S},~ \text{Tr}(W\rho_s) &\geq& 0, 
\end{eqnarray} 
with $\cal S$ being the set of separable states. The existence of such an operator is a consequence of the Hahn-Banach theorem on normed linear spaces \cite{simmons-book}. 
For every entangled state, there exists an entanglement witness. Note, however, given an entangled state, finding an optimal witness operator may not be an easy task~\cite{wit1,wit2,wit3}. 

In practice, if entanglement is required as a resource for a chosen information processing task, it is a particular entangled state that is aimed at, for implementing the task. To confirm the entanglement present in such a state, one is usually interested in performing the detection process using local measurements. Suppose the state that required in an information processing task is the two-party state, $|\psi^+\rangle\langle\psi^+|$, where $|\psi^+\rangle = \frac{1}{\sqrt{2}}(|01\rangle +|10\rangle)$. The preparation procedure may infuse some noise, and resultant state shared between the two parties may turn out to be
\begin{equation}
\label{mixed}
\rho=p |\psi^+\rangle\langle\psi^+| + (1-p)\sigma,
\end{equation}
where $\sigma$ is a two-qubit density matrix and $p$ is such that $\rho$ is positive semi-definite. Here, $\sigma$ represents the noise infusion, and $1-p$ represents the strength of the noise. Suppose that $||\sigma - \frac{1}{4}\mathbb{I} \otimes \mathbb{I} ||\leq d$, where $d \geq 0$, and $\mathbb{I}$ is the identity operator on the qubit Hilbert space. If $d=0$, then the noise is said to be ``white", but, in general, $d$ may not be zero. A witness operator that confirms the entanglement in   $|\psi^+\rangle$ reads as \cite{ew}
\begin{equation}
\label{wo}
W_0=|\phi^+\rangle\langle\phi^+|^{T_A}=\frac{1}{4}(\mathbb{I}\otimes\mathbb{I}+\sigma_z\otimes\sigma_z -\sigma_x\otimes\sigma_x -\sigma_y\otimes\sigma_y).
\end{equation}
It was shown that $W_0$ is also ``optimal" for $|\psi^+\rangle$, in the sense that $\langle \psi^+|W_0|\psi^+\rangle=\min_{W \in \mathbb{M}} \langle \psi^+|W|\psi^+\rangle $, where $\mathbb{M}$ is the collection of all witnesses for states on $\mathbb{C}^2 \otimes \mathbb{C}^2$ \cite{Park-10}. The witness $W_0$ remains optimal for the state $\rho$ in Eq.~(\ref{mixed}), provided $d=0$ \cite{ew}.
The advantage of this witness operator is that to implement it in a laboratory,  the observers, who may be spatially separated, have to perform three correlated local measurements in the bases corresponding to the Pauli operators $\{\sigma_x, \sigma_y, \sigma_z\}$.

It may be noted that entanglement witnesses are not only used for the detection of entanglement, but also for its quantification. It was shown in \cite{qew} that any measured negative expectation value of a witness can be turned into a non-trivial lower bound on generic entanglement measures. See also \cite{Eisert07,werner08-Reimpell}.

\subsection{Unsharp measurements}
 The quantum theory of measurement is counter-classical in the sense that in order to obtain information about the state, disturbance of the state becomes unavoidable, unless the state is diagonal in a measurement basis.
A von Neumann type measurement \cite{vn}, dubbed as ``strong" measurement, transforms the initial state of the system into one of the eigenstates of the measured
observable, assuming the measurement to be of rank-$1$ and repeatable. This type of measurement typically yields a large amount of information about the measured system, and leads to output states about which we have the maximum information that is quantum mechanically accessible. See \cite{fuchs97} in this regard.
On the other hand, there exist measurement schemes, such as weak measurements \cite{wk}, which provide less information about the system while affecting it only weakly.
It is important to mention here that we consider weak measurements without the associated pre- and post-selection procedures. More specifically, we employ the unsharp measurement formalism, which are a special subset of general positive-operator valued measurements (POVMs)~\cite{busch3}. In a practical situation, e.g. in a laboratory, measurements are almost always imprecise. This means that, for example, for a spin measurement, the pointer states of the apparatus corresponding to orthogonal spin states are not perfectly distinguishable. There is, therefore, the possibility of a non-zero overlap between such pointer states. This fuzziness of the apparatus states is captured by an unsharp measurement. It is to be noted that the terminology that we are using here identifies non-orthogonal POVM elements with pointer states that are not completely distinguishable. It is also possible to consider distinguishable pointers in a larger Hilbert space, via the Naimark theorem \cite{riesz-functional,peres-book}. For two-outcome measurements on the quantum spin-$1/2$ space,
the notion of unsharp measurement can be captured by the operator, $E_{\pm|\hat{n}}^{\lambda}=(\mathbb{I} \pm \lambda \hat{n}.\vec{\sigma})/2$, where $\vec{\sigma}=(\sigma_x,\sigma_y,\sigma_z)$, and $\hat{n}$ is a three-dimensional unit vector, with $\lambda\in(0,1]$~\cite{busch2}. 
Here $\lambda$ plays the role of the parameter that quantifies ``sharpness" of the measurement. Indeed for $\lambda=1$, $E_{\pm|\hat{n}}^{\lambda}$ correspond to projectors. Note that $E_{+|\hat{n}}^{\lambda}$ and $E_{-|\hat{n}}^{\lambda}$ are positive operators that add up to the unit operator.  Thus the set of effects $E_{\hat{n}}^{\lambda}=\{E_{+|\hat{n}}^{\lambda},E_{-|\hat{n}}^{\lambda}\}$ constitute a POVM. It is interesting to know that the elements of the POVM
can be written as linear combinations of sharp projectors with white noise:
\begin{eqnarray}\label{povm}
E_{\pm|\hat{n}}^{\lambda}
=\lambda P_{\hat{n}}^{\pm}+\frac{1-\lambda}{2}\mathbb{I}.
\end{eqnarray}
Here, $P_{\hat{n}}^{\pm}$ are the projectors corresponding to the sharp measurement of a quantum spin-$1/2$ system in the direction $\hat{n}$, so that $P_{\hat{n}}^{\pm}$ are projectors of eigenvectors of $\hat{n}.\vec{\sigma}$. Unsharp measurements have variously been referred in the literature as fuzzy, imprecise, or weak measurements \cite{busch3,busch-anotherbook}. 

{\bf {Rule for determining post-measurement state:}}
In our subsequent analysis, the state of the system after  performing the measurements is required in order to evaluate the statistics of the sequential measurements. 
Under unsharp measurements, the post-measured state is given, within the generalized 
von Neumann-L\"uders transformation rule~\cite{busch1}, upto unitary freedom, as
\begin{eqnarray}
\rho\rightarrow \frac{1}{\tilde{p}}\sqrt{E_{\pm|\hat{n}}^{\lambda}}\rho\sqrt{E_{\pm|\hat{n}}^{\lambda}},
\end{eqnarray} 
with probability $\tilde{p}=\mbox{Tr}\left(\sqrt{E_{\pm|\hat{n}}^{\lambda}}\rho\sqrt{E_{\pm|\hat{n}}^{\lambda}}\right)$.
This transformation rule generalizes the projection postulate of sharp measurements. 

\section{THE SCENARIO}
\label{scenarip}

\begin{figure}[t!]
    \begin{center}
    \resizebox{8cm}{5cm}
    {\includegraphics{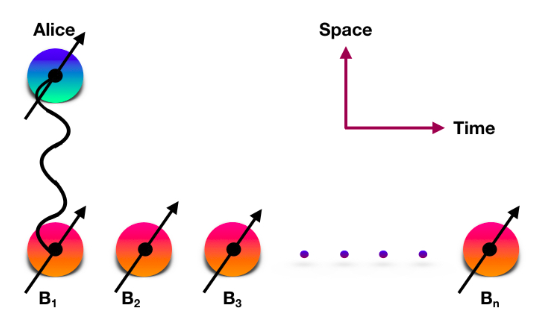}}
    \end{center}
\caption{(Color online.) Different Bobs, $B_1, ..., B_n,$ appear at the same scene (laboratory) to perform measurements on the same quantum particle on the Bob part of the Alice : Bob partition. 
The laboratory of Alice is spatially separated from that of the Bobs. In the schematic diagram, time-separation is depicted along the horizontal axis, while space-separation is represented along the vertical one.} 
    \label{lgif}
  \end{figure}
  
Let us now describe the scenario in which we work in this paper, for the distribution of the entanglement in the resource state, and
the corresponding arrangement in the laboratories hosting the state. A two-qubit entangled state is initially shared between two parties. One of the qubits is possessed by Alice, who always performs projective measurements, while the other qubit is possessed by $n$ Bobs, say, $B_1, B_2,\ldots, B_n$, who measure \emph{sequentially and independently}. See Fig.~\ref{lgif}. 
We now briefly describe the operational implications for the conditions of `sequentiality'' and ``independence" of the measurement strategy.

\emph{Sequentiality.}--
 The first Bob measures weakly with sharpness parameter $\lambda_1$. After $B_1$'s measurement, the qubit comes into possession of the second Bob, $B_2$, who measures on it with sharpness parameter 
$\lambda_2$. Similarly, the other Bobs, viz. $B_3, B_4, \ldots, B_{n-1}$, perform their measurements when they get the particles, with their corresponding sharpness parameters, determined by their apparatuses, except the last Bob, $B_n$, who measures sharply, i.e. with a unit sharpness parameter, so that the corresponding measurement is projection-valued. Such scenario can occur when either after measurement, each Bob sends his measured state via a noiseless channel to the next Bob or $B_1, B_2,\ldots, B_n$ perform measurements in the same laboratory but in different times. In each step, Alice-Bob examine whether the state is entangled or not. 

\emph{Independence.}--
We adopt the scenario where the Bobs measure independently,
which means that none of the Bobs are aware of the measurement settings of the others and hence the choice of a Bob's measurement, say, $B_i$, does not depend on the choices of previous measurements performed on second particle by $B_1, B_2, \ldots, B_{i-1}$.
The state possessed by a certain Bob is obtained by averaging over all the measurements and outcomes performed by all the previous Bobs. 

It is important to stress here that the ordering of the measurement performed by Alice and the measurements of the Bobs is not important because the measurement of Alice commutes with the measurements performed by Bobs. However, the ordering between the measurements performed by the Bobs is significant.  For the purpose of the treatment of the problem, we will assume that Alice performs her sharp measurement after the measurements of all the Bobs have been completed.

Let us now discuss 
about the modification of the witness operator (see Eq.~(\ref{wo})) which needs to be affected due to the fact that \emph{unsharp} measurements are being performed by the Bobs. 

\subsection{Modification of the witness operator due to unsharp measurements}
The joint probabilities for the shared state due to a sharp (projection) measurement by Alice and an unsharp measurement by one of the Bobs, in an intermediate stage of the measurement process, is of the form
\begin{equation}
\mbox{Tr}\Big(\rho(P^i_{\hat{n}} \otimes E_{j|\hat{m}}^{\lambda})\Big),
\end{equation}  
where $\rho$ is the average output state from the previous stage of the measuring process, $i,~j=\pm$, $P^i_{\hat{n}}$ is a projection operator corresponding to the projection measurement by Alice, and 
$E_{j|\hat{m}}^{\lambda}$ is a POVM element corresponding to the POVM by the Bob of this stage. The expectation value in the state $\rho$, corresponding to this joint measurement is given by 
\begin{equation}
\mbox{Tr}\Big[(P^+_{\hat{n}}-P^-_{\hat{n}}) \otimes (E_{+|\hat{m}}^\lambda - E_{-|\hat{m}}^\lambda)  \rho \Big].
\end{equation}
Note that $P^+_{\hat{n}}-P^-_{\hat{n}}$ is just $\hat{n}.\vec{\sigma}$. Let us denote it as $\sigma_{\hat{n}}$. Let us also denote $E_{+|\hat{m}}^\lambda - E_{-|\hat{m}}^\lambda$ as $\sigma^\lambda_{\hat{m}}$. Then, we have
\begin{eqnarray}
\label{use1}
\langle \sigma_{\hat{n}} \otimes \sigma_{\hat{m}}^\lambda \rangle \equiv \mbox{Tr}\Big[(P^+_{\hat{n}}-P^-_{\hat{n}}) \otimes (E_{+|\hat{m}}^\lambda - E_{-|\hat{m}}^\lambda)  \rho \Big] \nonumber\\
=\mbox{Tr}\Big[(P^+_{\hat{n}}-P^-_{\hat{n}}) \otimes \lambda (P_{\hat{m}}^+ - P_{\hat{m}}^-)  \rho \Big] \nonumber\\
=\lambda \langle \sigma_{\hat{n}} \otimes \sigma_{\hat{m}} \rangle.
\end{eqnarray}
Noting this relation and remembering that $W_0=\frac{1}{4}(\mathbb{I}\otimes\mathbb{I}+\sigma_z\otimes\sigma_z -\sigma_x\otimes\sigma_x -\sigma_y\otimes\sigma_y)$ was used (see Eq.~(\ref{wo})) as the witness for the state $|\psi^+ \rangle \langle \psi^+|$, when $\lambda=1$, 
we propose the substitution $\langle \sigma_{\hat{n}} \otimes \sigma_{\hat{m}} \rangle \rightarrow \lambda \langle \sigma_{\hat{n}} \otimes \sigma_{\hat{m}} \rangle$, in the case of a general $\lambda$, so that  the effective entanglement witness in this case becomes 
\begin{eqnarray}
\label{wow}
W_0^{\lambda}=\frac{1}{4} \Big(\mathbb{I}\otimes\mathbb{I}+\sigma_z\otimes\lambda\sigma_z -\sigma_x\otimes\lambda\sigma_x -\sigma_y\otimes\lambda\sigma_y \Big).\nonumber\\
\end{eqnarray}
It is easy to check that $\mbox{Tr}(W_0^\lambda \rho_s) \geq 0$ for all separable states $\rho_s$.

\section{SHARING OF ENTANGLEMENT BY MULTIPLE BOBS}
\label{sharing}

\subsection{Maximally entangled initial state}
Suppose that the maximally  entangled pure state $|\psi^+\rangle$ is shared between two spatially separated laboratories. An entanglement witness for this state is given by $W_0=|\phi^+\rangle\langle\phi^+|^{T_A}$ (see Eq.~(\ref{wo})). 

Corresponding to the measurement by Alice and $B_1$, the entanglement witness $W_0^{\lambda_1}$ acquires the expectation value 
\begin{equation}
\mbox{Tr}\big[|\psi^+\rangle \langle \psi^+| W_0^{\lambda_1}\big]=\frac{1}{4}(1-3\lambda_1).
\end{equation}
It is clear from the above expression that $\lambda_1>1/3$ is required for detecting entanglement by $B_1$, using the witness operator $W_0^{\lambda_1}$. Note that this value is lower than the threshold value of sharpness parameter required to demonstrate violation of Bell's inequality (which requires $\lambda_1 > \frac{1}{\sqrt{2}}$) \cite{sygp,sm}. This difference between the thresholds of the violation of Bell's inequality and entanglement detection may be expected as violation of local realism has been argued to require ``stronger" quantum correlations than just entanglement.
In particular, Bell inequalities, typically, form non-optimal witnesses \cite{ew,wernerstate}. 
Such an argument was possibly first put formed in 1989 by using the Werner state \cite{wernerstate},
i.e. the state in Eq.~(\ref{mixed}) for $d=0$, which is entangled for $\frac{1}{3} < p \leq 1$, while it violates Bell inequality only for $\frac{1}{\sqrt2} < p \leq 1$. 

Let us now explore if there is the possibility for subsequent observers at the laboratory of $B_1$, viz. $B_2$, $B_3$, $\ldots$, to share residual entanglements with Alice that can be detected through entanglement witnesses. Note that the possibility for this to happen has been created because of the fact that $B_1$ has performed an unsharp measurement. Sharp measurements by both Alice and $B_1$ would have resulted in a product state between the two laboratories. Note that we are considering only rank-$1$  measurements here, in the case of sharp  (projection) measurements. Note also, and this we have discussed in Sec.~\ref{scenarip}, that Alice's sharp measurement does not preclude  $B_2$'s ability to share entanglement with Alice.

As all the Bobs are ignorant about what measurements were performed by previous Bobs in a given run of experiment, we have to average over the previous Bob's input and output to obtain the state shared between Alice and the Bob of the current stage of the experiment. After performance of $B_1$'s unsharp measurement, the average state is given by
\begin{eqnarray}\label{avst}
|\psi^+\rangle \langle \psi^+| \to \rho_1^{\lambda_1} 
=\frac{1}{3} \sum_{i, \hat{n}} \sqrt{E_{i|\hat{n}}^{\lambda_1}} |\psi^+\rangle \langle \psi^+| \sqrt{E_{i|\hat{n}}^{\lambda_1}},
\end{eqnarray}
where \(i=\pm\), \(\hat{n}=\hat{x},\hat{y},\hat{z}\).
After some algebra, we obtain
\begin{eqnarray}
\rho_1^{\lambda_1}= \frac{1}{4} \Big[p \rho_{\psi^+}+(1-p) \mathbb{I} \otimes \mathbb{I}\Big],
\end{eqnarray}
where $p=\frac{1}{3}(1+2\sqrt{1-\lambda_1^2})$. 

In the next stage of the protocol, $B_2$ measures unsharply on his part of $\rho_1^{\lambda_1}$ with sharpness parameter $\lambda_2$, to check with Alice as to whether the state is entangled, by using the entanglement witness $W_0^{\lambda_2}$. The reason for using the same form of the entanglement witness as in the first stage (when $B_1$ is operating) is because the state shared by Alice and $B_2$, before their measurements, is in the Werner form \cite{ew} and $W_0^{\lambda_2}$ is an optimal EW operator for $\rho_1^{\lambda_1}$. 
With this state and these measurements, one obtains
\begin{eqnarray}
\mbox{Tr}[W_0^{\lambda_2} \rho_1^{\lambda_1}]=-\frac{1}{4} \Big[1-(1+2\sqrt{1-\lambda_1^2})\lambda_2 \Big].
\end{eqnarray}
 Now if $\lambda_1=1/3$ in the first stage, then to detect entanglement in the second stage, the sharpness parameter $\lambda_2$, of $B_2$, must be greater than $0.3465$ (correct up to four significant figures). This implies that  $B_2$ has to measure with more precision than $B_1$ to detect entanglement. If both  $B_1$ and  $B_2$ are to detect entanglement in their respective stages, then we must have $\lambda_1=\frac{1}{3}+\epsilon_1$, with $\epsilon_1>0$ (but $\epsilon_1 \leq \frac{2}{3}$), and we must correspondingly choose a $\lambda_2$ for  $B_2$, 
 so that $-\frac{1}{4}\big[1-\big(1+2\sqrt{1-(\frac{1}{3}+\epsilon_1)^2}\big)\lambda_2\big] < 0$.

Now in order to obtain the limit on the number of Bobs who can detect entanglement with a single Alice, we adopt the following procedure. In a similar way as described above,  $B_3$  measures on the average state obtained after measurements of  $B_1$ and  $B_2$. There is also a threshold value of 
$\lambda_3$ which is greater than $\lambda_1,~\lambda_2$. In general, for $n$ number of Bobs, one can find the condition of detection of entanglement by all the subsequent Bobs. The corresponding threshold values would be increasing, i.e., $\lambda_1<\lambda_2<\ldots<\lambda_n$. This process of choosing further Bobs can continue, with each Bob being able to detect entanglement in the average shared state obtained from the previous stage, as long as  $\lambda_n\ngtr 1$. From this condition, one can find the maximum number of Bobs sharing entanglement with a single Alice so that the shared entanglement can be detected through EWs. 

For the maximally entangled state, $|\psi^+\rangle$, shared initially between Alice and  $B_1$, 
 we find that $n=12$ i.e., at most twelve Bobs, acting sequentially, can detect entanglement with a single Alice. The bound on the number of Bobs in this case is significantly larger than the number of Bobs who can demonstrate violation of the CHSH inequality. Note that the average state becomes separable after twelve Bobs have performed sequential measurements with threshold values of the sharpness parameters. 
%

\subsection{Non-maximally entangled pure initial state and an operational entanglement measure: Singlet is not special}
In the preceding subsection, we found the limit on the number of observers witnessing entanglement with single Alice for a maximally entangled initial state. Now one may ask: if the initial entanglement is not maximal, but all other situations remaining the same, how many Bobs can detect entanglement with Alice? 
We restrict to pure shared states, and then the von Neumann entropy of the local density matrix is a 
good measure of entanglement \cite{Schumacher996}. If the number of Bobs scales with entanglement of the initially shared state, then we can have an operational measure of entanglement, via this corridor. We find that this is exactly the case, albeit in a coarse-grained form. On the way, we also find that the maximum number of Bobs who can detect entanglement, that is initially pure, with Alice remains unchanged for a finite range of near-maximal local von Neumann entropy. It may be noted that any two-qubit state with maximal local von Neumann entropy is local unitarily equivalent to the singlet state, $\frac{1}{\sqrt{2}} (|01\rangle-|10\rangle)$.  

A pure bipartite state can always be written, upto local unitaries, in the form $|\psi\rangle=a|01\rangle + b|10\rangle$, where $a$ and $b$ are real, and $a^2+b^2=1$. The entanglement content of this state can be quantified by the local von Neumann entropy $H(a^2)=-a^2 \log_2 a^2-b^2 \log_2 b^2$. For this state, the optimal entanglement witness remains the same as before, i.e., it is $W_0$ \cite{ew}. Suppose now that  $B_1$ measures weakly, with the sharpness parameter $\lambda_1$. Correspondingly, the expectation value of $W_0^{\lambda_1}$ is given by 
\begin{equation}
E_1=\frac{1}{4} [1 - (1 + 4ab) \lambda_1].
\end{equation}
Similarly, for the case when  $B_1$ and  $B_2$ measure weakly with sharpness parameters $\lambda_1$ and $\lambda_2$ respectively, we get
\begin{eqnarray}
E_2 &=& \mbox{Tr}[W_0^{\lambda_2} \rho_a^{\lambda_1}] \nonumber\\
&=& \frac{1}{4} \Big[1 - \frac{1}{3}(1 + 4ab) (1+2\sqrt{1-\lambda_1^2}) \lambda_2 \Big],
\end{eqnarray}
where $\rho_a^{\lambda_1}$ is the average state after $B_1$ performs 
his measurement on $|\psi\rangle$. Note that $\rho_a^{\lambda_1}=\rho_1^{\lambda_1}$ for $a=b=\frac{1}{\sqrt{2}}$.
Here it should be mentioned that unlike the case of maximally entangled initial state, here, the average state after weak measurement performed by a Bob becomes a mixed entangled state with colored noise. 
Even for this entangled state, $W_0$ remains a useful entanglement witness \cite{ew}, although it is non-optimal. We however continue to use the entanglement witness, $W_0$, which is optimal for any state in the class, $a|00\rangle+b|11\rangle$.
For $n$ Bobs measuring sequentially and independently, in the same way as in the preceding subsection, generalizing the above results, we find that
\begin{eqnarray}
E_n=\frac{1}{4} \Big[3^{n-1} - \frac{1}{3^{n-1}} (1 + 4ab) \lambda_n \Pi_{i=1}^{n-1} (1+2\sqrt{1-\lambda_i^2} ) \Big], \nonumber\\
\end{eqnarray}
where $n=1,2,3, \ldots$.

\begin{figure}[t]
\includegraphics[angle=0, width=85mm]{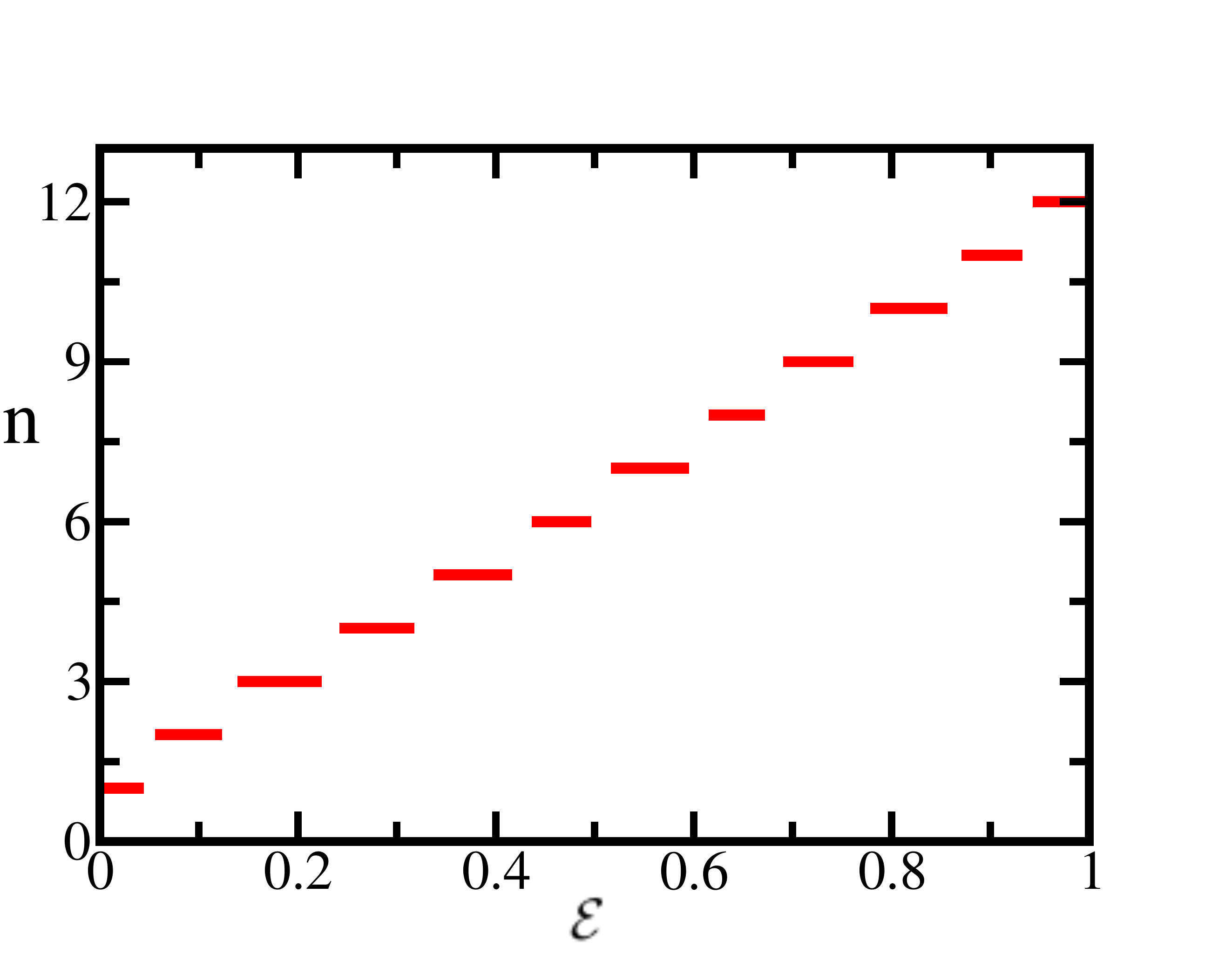}
\caption{(Color online.) 
Conceptualizing a coarse-grained operational entanglement measure. We consider the scenario where the two separated laboratories share a pure two-qubit state. The horizontal axis represents the entanglement of the initial state, as measured by the von Neumann entropy, $\cal{E}$, of one of the local states and is measured in ebits. The vertical axis counts the number of Bobs $(n)$ who can succeed in detecting entanglement with Alice, and is dimensionless. The monotonic nature of the function plotted implies that it can act as an entanglement measure, and it is clearly operationally defined. However, the steps in the function points to a coarse-grained nature of the measure. The existence of a  step of finite (i.e., non-zero) length on the extreme right implies that the maximal number of Bobs remains fixed for a certain range of ${\cal E}$, thereby indicating that pure states with  maximal local entropy like the singlet, do not have a special status in the scenario considered.
}
\label{fig_contour}
\end{figure}

We present our result in Fig.~\ref{fig_contour}, which indicates how many Bobs can detect entanglement with a single Alice, for a given pure initial shared state. It is clear from the figure that as the amount of local von Neumann entropy in the initial pure state decreases, the number of Bobs also decreases. It is also to be noted that except for a zero-measure set of values of local entropy, for initial states having amounts of local entropy close to each other, the number of successful Bobs remains the same. Specifically, Fig.~\ref{fig_contour} shows that for each $n$, there exists a continuous range of values of the local von Neumann entropy of the initial pure state, such that $n$ Bobs can detect entanglement with Alice. For example, we observe that at most twelve Bobs can detect entanglement with a single Alice if the local entropy of the pure initial state is more than $0.94$ ebits. Let us stress here that there are several situations known in the literature of quantum information where sharing a maximally entangled state is critical. Examples include quantum dense coding, both deterministic and probabilistic, and quantum teleportation \cite{dens,tele,barenco995,Hausladen996,mozes005, wu006, bourdon008, beran009, tsai010, song011}. On the contrary, the result obtained here is an addition to those cases where it has been shown that shared maximally entangled states do not have any exceptional stature \cite{huelga1997,acin002latorre,MAU2003}. It may be interesting to note that a coarse-grained measure of entanglement could still, in principle, provide a special status to the singlet (or any state that is local unitarily connected with the singlet) by providing its highest value only for such states. This is what happens, for example, in deterministic dense coding \cite{mozes005, wu006, bourdon008, beran009, tsai010, song011}, which has the same coarse-grained feature, but the maximal value is still reserved for the singlets or its local unitary cousins.
 Therefore, ``number of Bobs" defines a coarse but operational measure of entanglement. This measure can also be extended to   mixed entangled initial states. 

\subsection{Equivalent measurement devices for all Bobs}
\begin{figure}[t]
\includegraphics[angle=0, width=83mm]{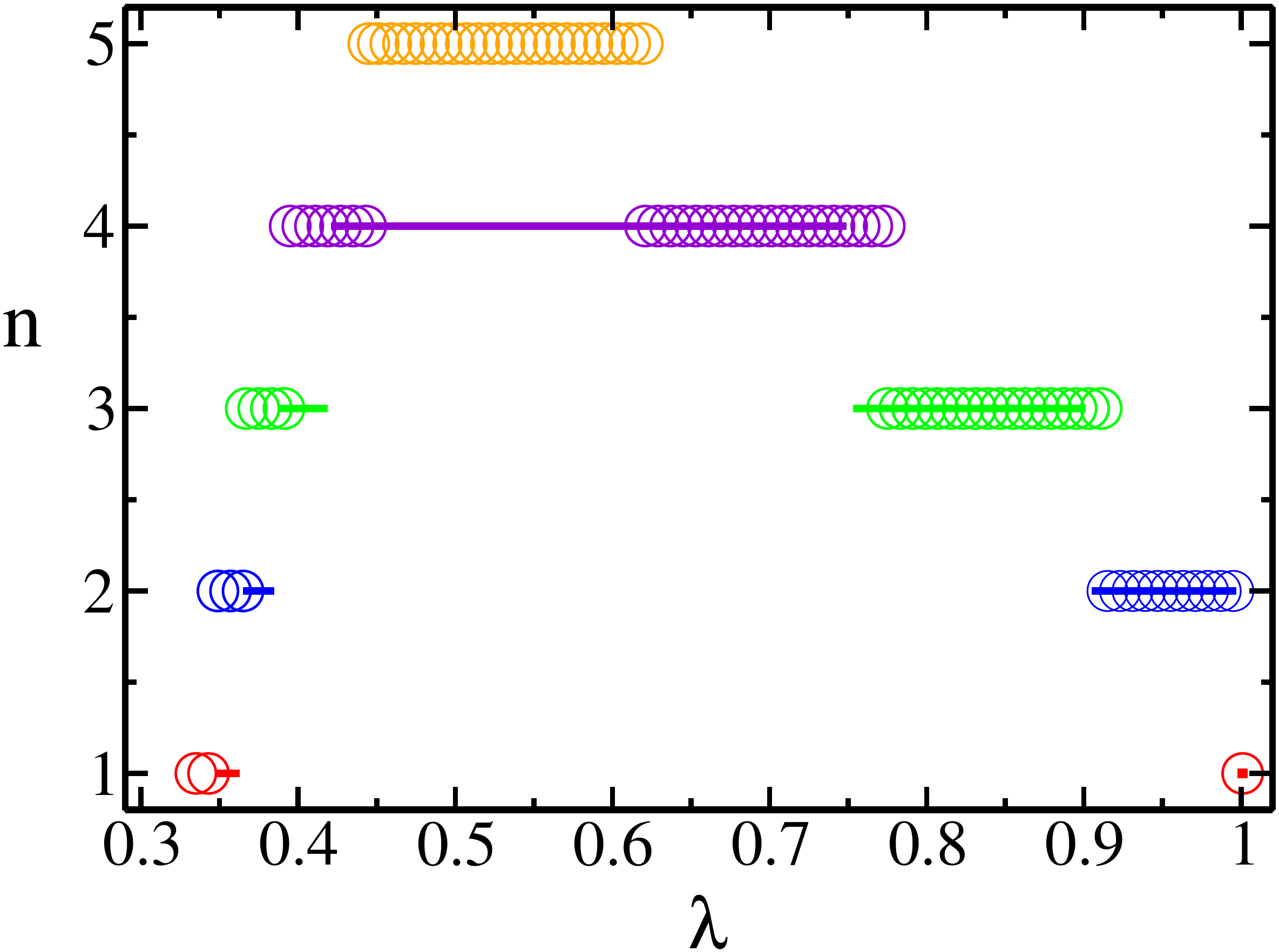}
\caption{(Color online.) 
The case of equal sharpness for all the observers measuring sequentially. The axis labelled $\lambda$ represents the common sharpness parameter of all the Bobs involved. The vertical axis stands for the maximal number of Bobs $(n)$ who are able to detect entanglement with a single Alice. Both axes represent dimensionless quantities. Circles and dashes exhibit the cases when the initial shared pure state is maximally entangled $({\cal E}(|\psi\rangle)=1)$ and when not so $({\cal E}(|\psi\rangle)=0.914)$, respectively. Red, blue, green, violet, and orange colors correspond to $n=1, 2, 3, 4, 5$, respectively.
}
\label{3dplot}
\end{figure}
We have until now, been working in the scenario where the sharpness parameters of the measurement apparatuses of the different Bobs could be different.
In this subsection, we consider a situation which in some instances can be more realistic than the one considered before. Precisely, Bobs are now constrained to use measurement devices with the same amount of sharpness. 
This means that the apparatus specifications are such that the associated sharpness parameters are the same for all the Bobs, i.e., comparing with the previous case, here we take $\lambda_1=\lambda_2=\ldots=\lambda_n=\lambda$ (say). In this case, it is clear from the previous result that the common sharpness parameter $\lambda$ should be $1/3$, or greater, so that at least one Bob can succeed in detecting entanglement. 
 
For a maximally entangled shared initial state,  we find that at most five Bobs can detect entanglement with a single Alice. 
In Fig.~\ref{3dplot}, we consider a maximally, and separately, a non-maximally entangled pure state, and provide the number of Bobs who can detect entanglement with Alice, under the restriction that all the Bobs use measuring apparatuses with the same value of sharpness, $\lambda$.
Interestingly, there arises an optimal range of the common sharpness parameter for which the number of Bobs is the highest. For a maximally entangled initially shared pure state, five Bobs can witness entanglement when $\lambda\in [0.45, 0.62]$ approximately. On the other hand, if the initial shared pure state is a non-maximally entangled, $|\psi\rangle$ with  $ {\cal E}(|\psi\rangle)\approx 0.918$, then the maximal number of Bobs who can detect entanglement is four, and this happens when $\lambda\in[0.42, 0.75]$ approximately. 
Just like in  the preceding subsection, we continue to use the witness $W_0$, for the state $a|00\rangle+b|11\rangle$ with colored noise, which is obtained after the second Bob has performed his weak measurement.
Note that for any given value of entanglement in the initial state, there is a specific value of the maximum number of Bobs who can detect entanglement with Alice, and this maximum is attained in a certain range of the sharpness parameter. As shown in the case of different measuring apparatuses, we also report here that five Bobs can detet entanglement not only for the maximally entangled initial state but for pure initial states with ${\cal E} \gtrsim 0.924$.

\subsection{Quantum discord of the final output state}  
We want to explore here whether there is any quantum correlation remaining after the last Bob's successful detection of entanglement with Alice. Such a quantum correlation has of course to be independent of entanglement. It is known that quantum discord \cite{discord1,discord2,modi-RMP,discord3} is a kind of quantum correlation which persists even in systems without entanglement. Let us consider the maximally entangled state $|\psi^+\rangle$ for which twelve Bobs measure on their part of the subsystem with threshold values of sharpness parameters. We find that the post-measurement averaged state, obtained after the twelfth Bob has performed his measurement, possesses a nonzero  quantum discord whose value is 0.0192 bits. It is interesting to note, therefore, that although there is no residual entanglement, in the post-measurement averaged state, some non-classical correlation persists, which can be quantified by quantum discord. 

\section{CONCLUSION}
\label{conclude}
Distribution of a resource state, like that of an entangled state, is the primary task of almost all  quantum information processing protocols. We investigated the problem of detection of entanglement when half of an entangled pair is possessed by an observer, called Alice, and the other half is sequentially and independently measured by several observers, called Bobs.

We found that for a maximally entangled shared state, at most twelve Bobs can detect entanglement with a single Alice, provided the measurements performed by the Bobs are weak or unsharp. The maximum number of Bobs remains invariant over a continuous range of  near-maximal entanglement -- up to $6\%$ lower than maximal -- in the initial pure shared state.
We also showed that the maximum number of Bobs decreases with decrease of entanglement content of the initially shared state, turning this number into an operational measure of entanglement. 
We observed that although there is no entanglement in the average state after the twelfth Bob has performed his measurement, the state still possesses quantum correlations in the form of quantum discord.

We also considered a more realistic scenario invoking the same sharpness parameter for measurement devices for all the Bobs. In this case, for maximal entanglement in the initial state, there is a range of the common sharpness parameter for which at most five Bobs can witness the entanglement with a single Alice. 
For any other value of entanglement in the initial pure state, there is an optimal number -- lower or equal to five -- of Bobs, and this optimality occurs in a particular range of the associated sharpness parameter. Again, the maximal number of Bobs remains unchanged for a continuous range of near-maximal initial pure entanglement.


\begin{center}
\textbf{Acknowledgments}
\end{center}
A.B. acknowledges the support of the Department of Science and Technology (DST), Government of India, through the award of an INSPIRE fellowship.

\end{document}